\documentclass[manuscript,screen,nonacm]{acmart}
\usepackage{xcolor}

\usepackage{booktabs}
\usepackage{enumitem}
\usepackage{longtable}
\usepackage{array}
\AtBeginDocument{%
  }

\usepackage{listings}
\lstdefinestyle{prompts}{
  basicstyle=\ttfamily\footnotesize,
  breaklines=true,
  breakatwhitespace=false,
  columns=fullflexible,
  keepspaces=true,
  frame=single,
  framerule=0.3pt,
  xleftmargin=0.5em,
  xrightmargin=0.5em,
  aboveskip=0.7em,
  belowskip=0.7em,
  showstringspaces=false,
  upquote=true
}
\begin{document}

\title{AI Persuasive Framing in Collective Dilemmas}

\author{Anders Giovanni Møller}
\authornote{Both authors contributed equally to this research.}
\email{andersgiovanni@gmail.com}
\orcid{0009-0008-9737-555X}
\affiliation{%
  \institution{Data Science Section, IT University of Copenhagen}
  \city{Copenhagen}
  \country{Denmark}
}

\author{Alessia Galdeman}
\authornotemark[1]
\email{alessia.galdeman@unimi.it}
\orcid{0000-0003-3286-4666}
\affiliation{%
  \institution{Computer Science Department, University of Milan}
  \city{Milan }
  \country{Italy}
}

\author{Arianna Pera}
\email{arianna.pera22@gmail.com}
\orcid{0009-0009-4316-3428}
\affiliation{%
  \institution{Copenhagen Center for Social Data Science, University of Copenhagen}
  \city{Copenhagen}
  \country{Denmark}}

\author{Luca Maria Aiello}
\email{luai@itu.dk}
\orcid{0000-0002-0654-2527}
\affiliation{%
  \institution{Data Science Section, IT University of Copenhagen}
  \city{Copenhagen}
  \country{Denmark}
}

\renewcommand{\shortauthors}{Møller et al.}

\begin{abstract}
AI agents are promising tools that can act as flexible behavioral nudges to enhance human cooperation in addressing large-scale societal problems. However, evidence on whether AI agents can effectively boost cooperation remains mixed. We recruited 1,283 participants to play iterated Collective Risk Games in small groups, testing whether AI assistants could nudge participants toward cooperation. By using persuasive framing personalized to each player's Social Value Orientation profile, the AI interventions significantly increased contributions and group success rates. These cooperative effects were short-lived, however, fading after the first few rounds. Strikingly, when the AI treatments were reconfigured to promote selfish behavior through exculpatory framing, the negative effects on contributions and group success were larger and substantially more persistent, particularly for personalized interventions. This asymmetry between prosocial and antisocial persuasion highlights the dual-use risks of AI systems designed to influence group behavior in collective action settings.
\end{abstract}

\maketitle

\section{Introduction}

The hope that AI can help build more robust and resilient communities has motivated a wealth of recent work on human-AI interaction~\cite{tsvetkova2024new}. A central thread of this research is the goal of enhancing human cooperation through large-scale AI interventions~\cite{mikhaylovskaya2024enhancing, tessler2024ai}, as nudging behavioral change toward cooperation at scale is becoming increasingly necessary to address the most pressing societal challenges of this century, chief among them the climate crisis~\cite{stern2025green,lee2023ipcc}. This goal feels ever more attainable as AI agents powered by Large Language Models (LLMs) grow more capable of autonomously managing complex social interactions~\cite{chen2026ai,gao2024large} and of producing persuasive arguments~\cite{breum2024persuasive,salvi2024conversational}.

Behavioral economics experiments built around game-theoretical scenarios provide ideal testing grounds for assessing how group outcomes are shaped by AI interventions of various kinds~\cite{koster2022human}. These experiments typically involve goal-oriented settings in which multiple participants engage in a series of repeated interactions~\cite{osborne1994course}, and they can be used to ask questions such as whether AI agents interacting with participants can lower critical synergy thresholds required for successful cooperation~\cite{hintze2026promoting}. Within this nascent field, the debate over the effectiveness of AI interventions remains very much open, with conflicting evidence on offer. Yet, a growing body of work supports the hypothesis that AI may have limited or no effect on boosting cooperation in collective games~\cite{brito2025assistant, renoux2025effect, shen2025impact, qiao2026norms, anwar2026playing}.

One possible explanation for these null results is that many experiments in this line of work inherit a behavioral economics perspective on cooperation games, which emphasizes the availability of information about the system as a primary driver of the evolution of cooperation~\cite{kleshnina2023effect}. This perspective has inspired the deployment of AI solutions aimed at resolving such information bottlenecks~\cite{qiao2026norms}, while engaging little with research on persuasive and personalized AI interventions~\cite{matz2024potential}, which is instead concerned with generating interactions that can effectively trigger behavioral change, independent of the surrounding information context~\cite{huang2023artificial}.

In this work, we help fill this gap by testing AI assistants that use persuasive and personalized framing to motivate participants in a public goods game to increase their pledges to the group. Specifically, we recruit 1,283 participants to play an iterated collective risk game in small groups, and we compare a control condition with three separate treatments designed to increase pledges: (i) a static prognostic message that acts as a nudge; (ii) an AI assistant instructed to generate persuasive frames promoting prosocial behavior; and (iii) a semi-personalized AI assistant that adapts its persuasive strategy to the social value orientation of each participant.

We find that the AI interventions can increase pledges and the success rate of the game significantly, and more substantially than a simple static nudge, with longer chat sessions with the AI being associated with stronger effects. Consistent with prior work, the interventions have a fleeting effect that fades after the first few rounds of the game. Concerningly however, when the AI treatments are reconfigured to nudge uncooperative behavior through exculpatory framing~\cite{benabou2018narratives}, the negative effect on pledges and game success is even larger and much more persistent, particularly for semi-personalized interventions. 

Our contribution provides evidence for the need to more systematically integrate the literatures on human-AI cooperation and persuasive AI, while highlighting the potential risks of AI's efficacy in disrupting positive group dynamics when it is designed to do so.

\section{Related Work}

\paragraph{Public Goods Games and the Collective Risk Dilemma}
Public goods games (PGGs) are a foundational paradigm for studying cooperation under social dilemmas, in which individually rational behavior leads to collectively suboptimal outcomes. In a standard PGG, group members decide how much to contribute to a shared pool that benefits everyone, creating a tension between personal gain and collective welfare~\cite{palfrey1984participation, ledyard1994public}. A large body of work has examined the conditions under which cooperation emerges, with key findings pointing to the roles of punishment, reciprocity, group size, and social norms~\cite{fehr2000cooperation, fehr2002strong, pereda2019group, szekely2021evidence}.

The Collective Risk Game (CRG) is a threshold-based PGG in which groups must collectively reach a contribution target to avert a shared loss, and failure becomes increasingly likely when individual incentives to free-ride go unchecked~\cite{milinski2008collective}. This framing maps naturally onto high-stakes real-world dilemmas such as climate change mitigation, where the costs of action are immediate and individual while the benefits are delayed and collective~\cite{milinski2008collective, zhao2023cooperation}. Research in this context has confirmed the sensitivity of collective outcomes to the level of risk, threshold design, and strategic uncertainty~\cite{abou2012evolutionary, szekely2021evidence}.
Within this framework, social norms have emerged as a central mechanism sustaining cooperation. Szekely et al.~\cite{szekely2021evidence} showed that both empirical expectations (what others will do) and normative expectations (what others think one should do) are strong positive predictors of contributions. Crucially, their longitudinal results suggest that norms consolidate over repeated game rounds, and that once stabilized, behavior becomes less sensitive to changes in risk. Related work on CRGs has similarly emphasized that feedback, timing, and repeated play can substantially shape the evolution of contributions over time~\cite{abou2012evolutionary, chen2012averting}. 
Structural features of interaction and personal traits also matter: richer communication channels generally improve coordination~\cite{bicchieri2007computer}, and individual differences can further modulate how people respond to the same game setting~\cite{gavrilets2024co}.

\paragraph{Framing, Personalization, and Cooperative Behavior}
A growing body of literature examines how targeted communication can shift cooperative behavior. In the context of climate change, explicitly framing environmental risks has been shown to increase engagement with mitigation activities~\cite{farjam2018does}. Within PGGs more broadly, inter-player communication of contribution intentions has been found to substantially improve coordination, particularly under conditions of inequality where reaching the collective target is otherwise difficult~\cite{tavoni2011inequality}.

More directly relevant to the present work is research on experimenter-generated message frames. Gavrilets et al.~\cite{gavrilets2024co} found that presenting players with a message indicating what the group-optimal action would have been significantly shifted subsequent behavior. Crucially, the direction of that shift depended on participants' psychological and cognitive profiles. This heterogeneity in response was also found by Tverskoi et al.~\cite{tverskoi2023disentangling}, who showed that the same group-norm message could either strengthen personal norms or increase conformist behavior, depending on players' social value orientation. Together, these findings suggest that the effectiveness of any given message is not uniform.
Work on linguistic alignment and interaction format further supports the idea that matching message style to individual characteristics can improve strategic outcomes \cite{agranov2024communication, matzinger2024inherent}. 
This sensitivity to individual characteristics creates a natural opening for AI-mediated persuasion, where messages can be dynamically adapted in ways static frames cannot.

\paragraph{AI-Mediated Persuasion}
Large Language Models (LLMs) have recently been studied as persuasive agents in cooperative and social contexts. 
Breum et al. \cite{breum2024persuasive} employed LLMs to generate persuasive arguments in a climate change dialogue between agents, finding that when humans rated the resulting arguments, knowledge-based ones were judged most persuasive.
Personalization substantially amplifies this influence: Matz et al. \cite{matz2024potential} showed that tailoring messages to individual profiles improves persuasive impact, and Salvi et al. \cite{salvi2024conversational} found that personalized LLM agents can outperform human opponents in debates. 
Karinshak et al. \cite{karinshak2023working} further showed that personalized AI-generated messages can be perceived as more authoritative than official sources.
Beyond immediate persuasion, Costello et al. \cite{costello2024durably} demonstrated that sustained, personalized AI conversations can durably reduce belief in conspiracy theories, suggesting that such interventions may produce lasting prosocial effects. 
As models continue to grow in scale and capability, their persuasive influence appears to be increasing \cite{durmus2024measuring}, making the design of AI-mediated communication an increasingly relevant research question.

\vspace{0.5cm}
\noindent Prior work has established that cooperative behavior in threshold games is sensitive to framing and social norms, and that LLMs are increasingly capable persuasive agents. Yet these threads have not been brought together in a single experimental setting. The present study does so, situating an interactive AI agent within a collective risk game and testing personalization and direction of persuasion as orthogonal dimensions of influence.

\section{Methods}

\begin{figure}
    \centering
    \includegraphics[width=1\linewidth]{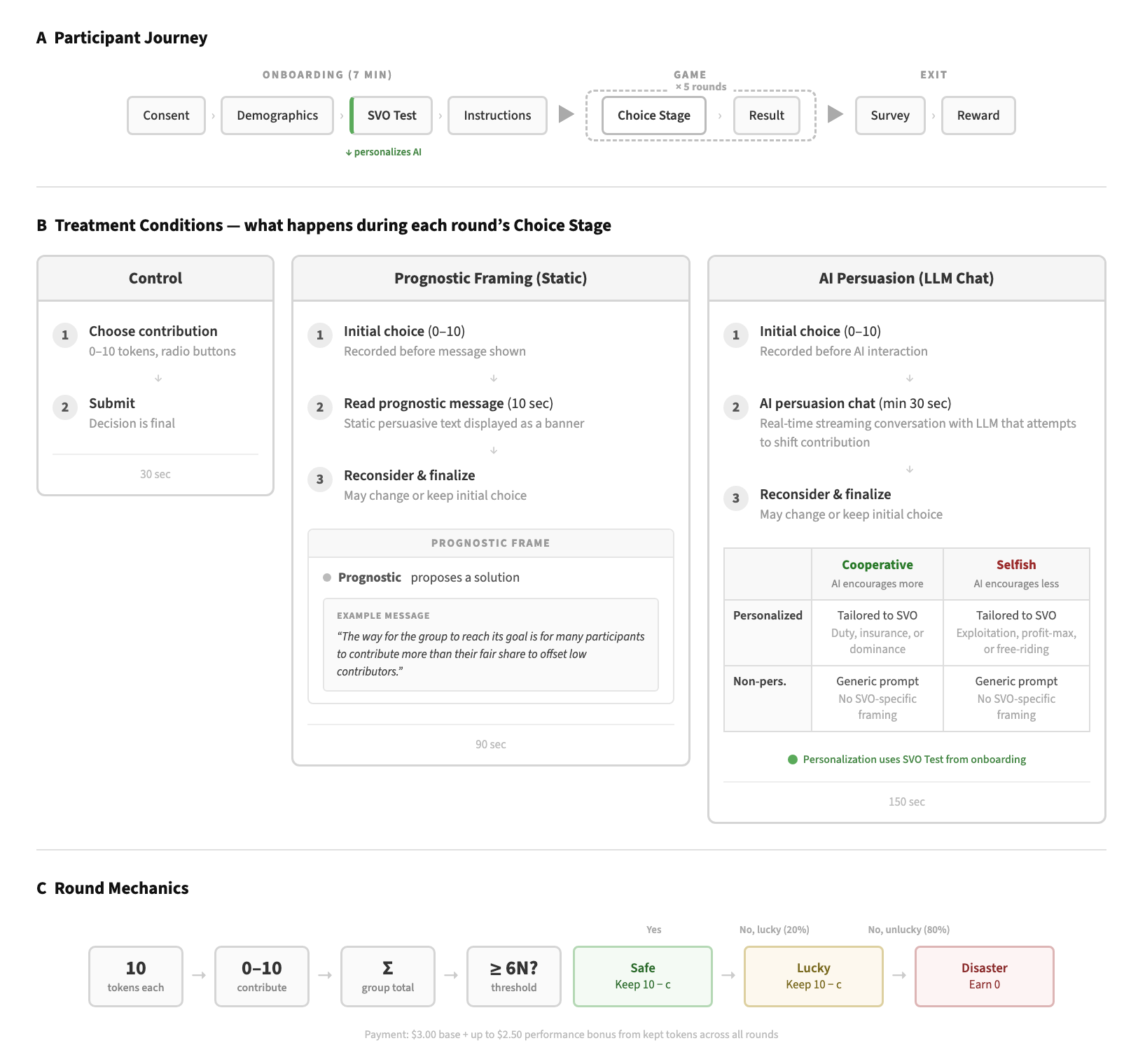}
    \caption{\textbf{Experimental Overview}. \textbf{(A)} The participant journey included onboarding, five game rounds, a post-study survey, and a reward page. \textbf{(B)} During each round, participants either submitted their contribution directly (control), reconsidered it after reading a static message, or interacted with an LLM encouraging either cooperative or selfish behavior. Messages and LLM interactions could be personalized based in information from the onboarding. \textbf{(C)} Participants received 10 tokens and contributed between 0 and 10 tokens. If the total contribution reached the 6N threshold, participants kept their remaining tokens. If not, they would loose them with 80\% probability. Code is available at \url{https://anonymous.4open.science/r/ai-in-collective-dilemmas-2617/}}
    \label{fig:experiment}
\end{figure}

\subsection{Collective Risk Game}
The experiment centers on the \textit{Collective Risk Game}, a public goods dilemma in which groups of participants must collectively contribute a shared resource to prevent a common disaster. In our experiment, the disaster is a fictional flooding scenario. This paradigm creates a tension between individual self-interest, since participants keep any resources they do not contribute as personal earnings, and collective benefit, which requires contributing enough to prevent the disaster. If the disaster is prevented, participants keep their remaining resources; otherwise, those resources are lost.

The game follows a five-round structure. In each round, every participant receives 10 fictional resource tokens and can individually contribute between 0 and 10 of them. The collective group threshold required to thwarth the disaster is defined as $6 \times \text{number of players}$, and the total group contribution is the sum of all active players' contributions. If the total equals or exceeds the threshold, the disaster is prevented and players keep their unspent tokens. If the total falls below the threshold, there are two possible outcomes. With $80\%$ probability, the disaster occurs and all players earn 0 tokens for that round. With $20\%$ probability, the players are lucky and no disaster happens, allowing them to keep their unspent tokens despite missing the threshold.

We recruit participants through the crowdsourcing platform Prolific~\cite{palan2018prolific}. Each participant receives a base payment of \$3 plus a bonus of up to \$2.5, depending on how they play the game. The round bonus is computed as $\text{round bonus} = \frac{\text{tokens kept}}{10}\times\frac{\text{max bonus}}{\text{number of rounds}}$. This amounts to \$0.5 per round if all 10 tokens are kept and the disaster is avoided. We set up rooms with 5 players, dynamically adjusting the minimum contribution threshold across rounds in the event of participant dropout, and we discard a game from analysis if the number of participants falls below 3.

\subsection{Experimental Design}

\subsubsection{Treatment conditions} 

Our experiment uses a between-subjects design, with groups randomly assigned to one of three conditions. The treatment is applied at the group level, so that all five participants are in the same condition.

\textbf{\textit{1. Control}}. In line with the game description, players make a single contribution in each round within a 30-second window. This establishes a baseline for contribution behavior in the absence of any intervention.

\textbf{\textit{2. Message Framing}}. Players first make an initial contribution choice. They then see a short message with prognostic framing that encourages cooperative solutions, consistent with the framing studied extensively in the consensus mobilization literature~\cite{benford2000framing} (see messages in Appendix~\ref{app:prognostic-messages}). Next, they are asked to confirm or revise their contribution. Five different variants of the message are shown across rounds in a randomized order. This condition tests whether a single, predefined persuasive cue is sufficient to shift behavior.

\textbf{\textit{3. AI Persuasion}}. Players first make an initial contribution choice, then engage in a real-time text conversation with an LLM-based AI agent for at least 30 seconds, and finally reconsider and finalize their choice. Unlike the static condition, the AI adapts its arguments to the player and their initial choice, producing a personalized dialogue rather than a fixed message. This condition tests whether interactive and responsive AI systems affect user behavior. All prompts are listed in Appendix~\ref{app:prompts}.

Conditions 2 and 3 share the same three-step structure: initial choice \textrightarrow \space influence \textrightarrow \space reconsider. We record both initial and final choices, allowing us to observe whether and how much players revise their contributions in response to the influence and across rounds.

The participant journey consists of three main stages: \textit{onboarding}, the \textit{game}, and the \textit{exit} phase. During the \textit{onboarding} phase, participants must accept a consent form, take a simple demographic survey, and complete a standard test to estimate their Social Value Orientation (SVO)~\cite{murphy2011measuring} before they are presented with the instructions of the game (see Appendix~\ref{app:svo-test} for SVO questions). The SVO test assesses people's preference for how to allocate resources between themselves and others. After the onboarding, they continue to the 5-round game phase where they must decide on their contribution followed by the result of the round. After the five rounds, they fill out an exit survey before they finally get a token for their reward.  

\subsubsection{Personalization}
We test whether personalizing the external influences has an additional impact on user behavior.

The AI assistant generates semi-personalized messages based on each participant's Social Value Orientation class, measured during the SVO test during onboarding. The scores from the different items are combined to classify participants as \textit{cooperative}, \textit{individualistic}, or \textit{competitive}. For cooperative users, the AI draws on arguments related to duty, insurance, and collective care. For individualistic users, the AI emphasizes maximization of personal gains. For competitive users, the AI suggests maximizing relative gains compared to the other players. In each round, the AI is prompted with information on the player's SVO type, their initial choice, and their contribution history. The non-personalized variant simply omits the SVO portion of the prompt.

As an orthogonal dimension of personalization, we also vary the \textit{direction of persuasion}. In the \textbf{cooperative} setting, the AI agent encourages high contributions. In the \textbf{selfish} setting, it encourages lower contributions. By including both directions, we can test not only whether personalized influence is more effective than generic influence, but also whether its effectiveness is symmetric across pro-social and anti-social objectives.

We also experimented with personalized messages in the static Message Framing condition. Drawing on the consensus mobilization literature~\cite{benford2000framing}, we considered five message types based on user profile: 1) \textit{diagnostic} frames that identify the nature of the problem; 2) \textit{prognostic} frames that propose a concrete solution; 3) \textit{motivational} frames that call users to action; 4) \textit{expert} frames that appeal to technical expertise; and 5) \textit{legitimate} frames that appeal to institutional authority. Based on pilot experiments, all frame types performed similarly, with \textit{prognostic} frames showing a slight advantage. In the interest of keeping crowdsourcing costs manageable, we therefore adopted a single static Message Framing condition using \textit{prognostic} frames exclusively.

\subsection{Analysis}

To study the impact of the treatments, we measure three key outcomes. Within each game, we compute the \textit{average decision}, defined as the mean contribution per round. We measure the \textit{success rate} as the per-round proportion of rounds in which the group collectively reached the threshold. In the treatment conditions, we also compute the per-round \textit{average decision change} as the mean difference between each player's final and initial pledges, averaged across all players. We report mean values across all games together with 95\% confidence intervals.

We compare each treatment against the control condition on \textit{average decision} and \textit{success rate}. For \textit{average decision}, we use the \textit{Mann-Whitney U} test, since each value is a discrete and bounded integer between 0 and 10. For \textit{success rate}, we use \textit{Fisher's exact test} on the $2 \times 2$ table of success and failure counts for control versus treatment.

In addition to these standalone metrics, we model whether the AI treatments actually affect contributions from one round to the next and from the initial pledge to the revised pledge within the same round. To do this, we fit OLS regression models with available covariates to predict three dependent variables:
\begin{align}
\text{contribution}_{r_{i}}
  &= \beta_{0}
   + \beta_1 \cdot 1[\text{SVO} = \text{cooperative}] \notag \\
  &\quad + \beta_2 \cdot 1[\text{SVO} = \text{individualistic}] \notag \\
  &\quad + \beta_3\,\text{success}_{r_{i-1}}
   + \beta_4\,\text{num\_players}_{r_{i}}
   + \beta_5\,\text{contribution}_{r_{i-1}} \notag \\
  &\quad + \beta_6\,\Delta_{\text{within-round}_i}
   + \beta_7\,\text{num\_interactions}_{r_{i}} \notag \\
  &\quad + \beta_8 \cdot 1[\text{treatment} = \text{personalized}] \notag \\
  &\quad + \beta_9 \cdot 1[\text{treatment} = \text{non\text{-}personalized}]
   + \varepsilon_i
  \label{eq:model-A} \\[4pt]
  \Delta_{\text{within-round}_i} \;\equiv\; \text{FinalPledge}_{r_i} - \text{InitialPledge}_{r_{i}}
  &= \text{same RHS as Eq.~\eqref{eq:model-A}, but excluding }\Delta_{\text{within-round}_i}.
  \label{eq:model-C}
\end{align}
The variable $\text{contribution}_{r_{i}}$ is the player's contribution in the current round and $\Delta_{\text{within-round}_i}$ is the difference between the final and initial pledge within a round. Among the independent variables, the SVO parameters indicate Social Value Orientation classes, with the reference category being unclassified, which applies when the SVO type could not be determined from the questionnaire (a player must provide at least six out of nine responses consistently within a type to be classified). In our cohort, no participants was classified as \textit{competitive}, leaving us with the two remaining classes of \textit{cooperative} and \textit{individualistic}. $\text{success}_{r_{i-1}}$ is a binary variable indicating whether the previous round reached the threshold, $\text{num\_players}_{r_{i-1}}$ is the group size after accounting for inactive or dropped-out participants, and $\text{contribution}_{r_{i-1}}$ is the player's contribution in the previous round. $\Delta_{\text{within-round}_i}$ is included as a predictor in equations~\eqref{eq:model-A}. The $\text{treatment}$ parameter reflects the assigned treatment condition, with control as the reference.

In addition to modeling round-to-round transitions, we fit analogous regression models for round 1 to better understand the factors influencing the first contribution and the within-round change between the initial pledge and the final decision in round 1. The variables are similar to those in~\eqref{eq:model-A}, but without any reference to prior rounds. For the control group, $\text{num\_interactions}_{r_{1}}$ and $\Delta_{\text{within-round}_1}$ are set to 0.
\begin{align}
\text{contribution}_{r_{1}}
  &= \beta_{0}
   + \beta_1 \cdot 1[\text{SVO} = \text{cooperative}] \notag \\
  &\quad + \beta_2 \cdot 1[\text{SVO} = \text{individualistic}] \notag \\
  &\quad + \beta_3\,\text{num\_players}_{r_{1}}
   + \beta_4\,\text{num\_interactions}_{r_{1}} \notag \\
  &\quad + \beta_5\,\Delta_{\text{within-round}_1} \notag \\
  &\quad + \beta_6 \cdot 1[\text{treatment} = \text{personalized}] \notag \\
  &\quad + \beta_7 \cdot 1[\text{treatment} = \text{non\text{-}personalized}]
   + \varepsilon_i
  \label{eq:model-A-prime} \\[4pt]
  \Delta_{\text{within-round}_1} \;\equiv\; \text{FinalPledge}_{r_1} - \text{InitialPledge}_{r_{1}}
  &= \text{same RHS as Eq.~\eqref{eq:model-A-prime}, but excluding }\Delta_{\text{within-round}_{r_1}}.
  \label{eq:model-C-prime}
\end{align}
\section{Results}

Our pool of 1,283 participants from Prolific is filtered for inactivity and low-quality responses distributed across 307 games (Control: 52, Prognostic: 51, Cooperative (pers.): 52, Cooperative (non-pers.): 50, Selfish (pers.): 52, Selfish (non-pers.): 50). Participants have a mean age of 40.5 years (see distribution in Figure~\ref{fig:age-distribution}) and a gender distribution of 50.7\% and 49.3\% for men and women respectively. In Table~\ref{tab:treatment-summary} we show key statistics across conditions.

\begin{table}[b!]
    \centering
    \begin{tabular}{lrrrr}
        \toprule
        Treatment & Players & Avg. Thresholds Met pr. Game & Avg. Contribution & Avg. Bonus \\
        \midrule
        Control                  & 241 & 3.75 & 6.30 & \$0.68 \\
        Prognostic               & 207 & 3.76 & 6.35 & \$0.70 \\
        Cooperative (Non-pers.)  & 217 & 4.29 & 6.62 & \$0.72 \\
        Cooperative (Pers.)              & 199 & 4.20 & 6.57 & \$0.71 \\
        Selfish (Non-pers.)      & 206 & 2.54 & 5.62 & \$0.54 \\
        Selfish (Pers.)                  & 213 & 2.04 & 5.41 & \$0.49 \\
        \bottomrule
    \end{tabular}
    \caption{Summary statistics by treatment. }
    \label{tab:treatment-summary}
\end{table}

\begin{figure}
    \centering
    \includegraphics[width=0.65\linewidth]{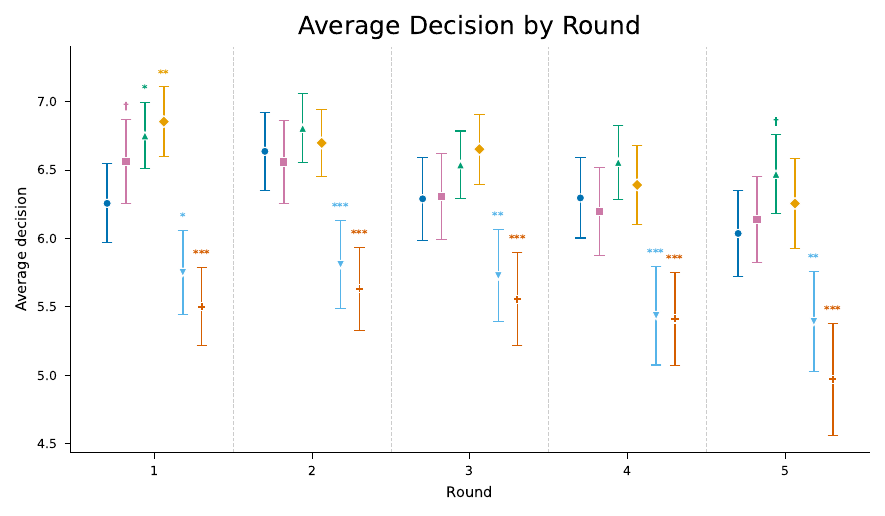}
    \includegraphics[width=0.65\linewidth]{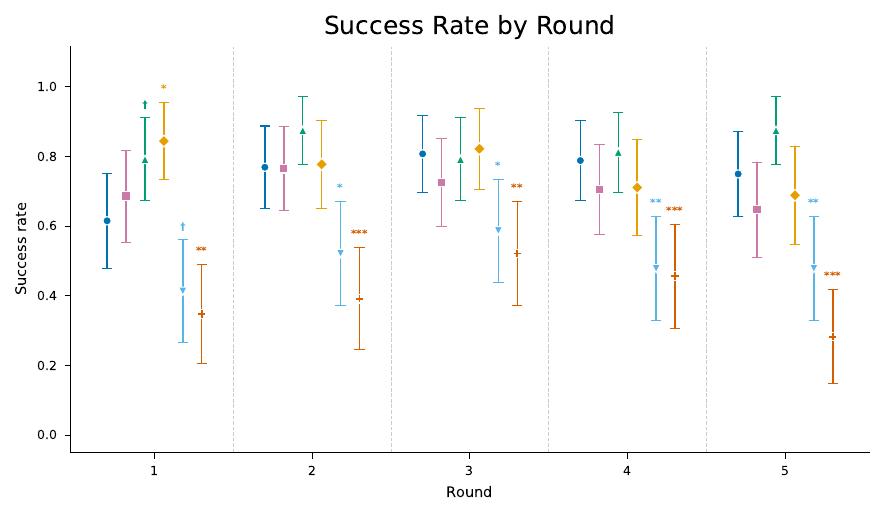}
    \includegraphics[width=0.65\linewidth]{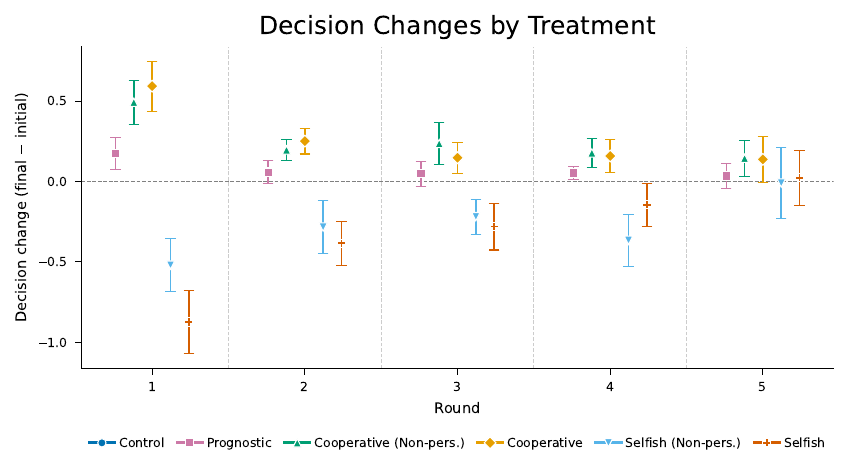}
    \caption{Effects of interventions across rounds. \textbf{Top:} Average contributions. \textbf{Middle:} Average group success rates. \textbf{Bottom:} Within-round changes from participants' initial pledge to their final contribution. Points show condition means and error bars indicate 95\% confidence intervals. Positive within-round changes indicate increased contribution, while negative changes indicate reductions. Asterisks reflect significant difference from control condition ($\dagger = p<0.10$, $* = p<0.05$, $** = p<0.01$, $*** = p<0.001$).}
    \label{fig:effects}
\end{figure}

\begin{figure}
    \centering
    \includegraphics[width=0.9\linewidth]{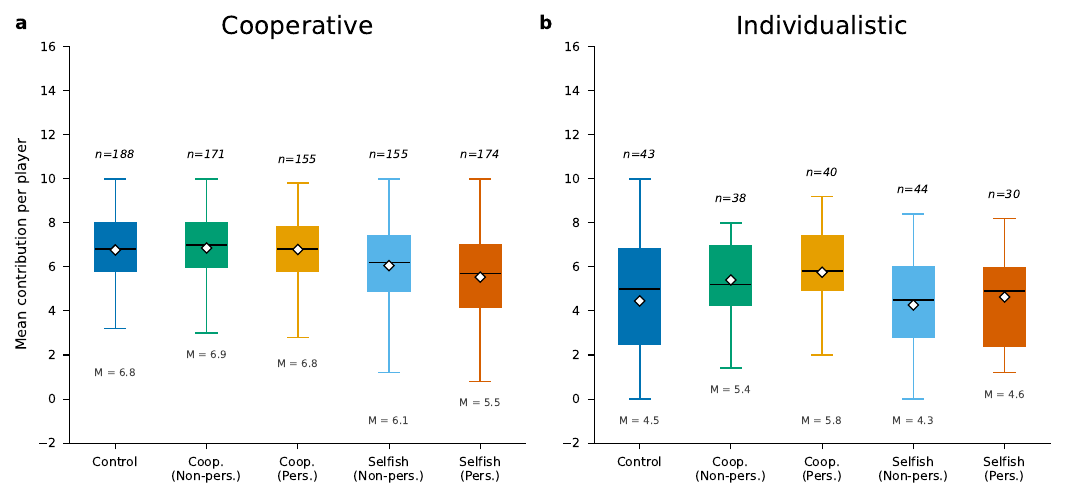}
    \caption{Distributions of mean player contribution across all games, disaggregated by treatment. \textbf{(a)} Distributions for players with a cooperative SVO profile. \textbf{(b)} Distributions for players with an individualistic SVO profile.}
    \label{fig:aggregates}
\end{figure}

Figure~\ref{fig:effects} shows how the average pledge, success rate, and within-round difference between initial and revised pledges vary across rounds. Focusing on the first round, we see that increasingly sophisticated interventions produce larger effects. Personalized AI interventions increase contributions and success rates the most, followed by non-personalized AI and then by static prognostic nudges. The selfish AI assistant shows the strongest and most significant effect, especially in its personalized version. Although interventions are delivered in every round, their effect appears to fade after round 1, with the exception of the anti-social nudges, whose effect remains significant throughout the game. In aggregate, all interventions yield similar average contributions per player, with a few caveats (Figure~\ref{fig:aggregates}). When disaggregating players by cooperative versus individualistic SVO profile, selfish AI is the only intervention that deviates from the control baseline for cooperative individuals, while the opposite holds for individualistic players, where cooperative personalized AI yields significantly higher contributions compared to the baseline.

\begin{figure}
    \centering
    \includegraphics[width=0.9\linewidth]{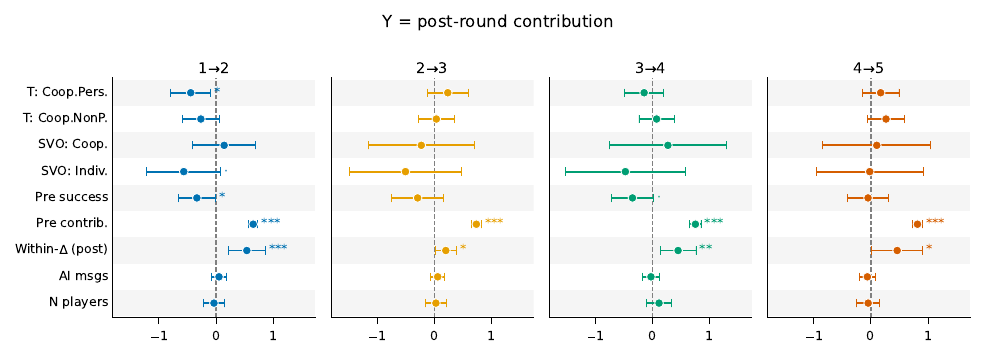}
    \includegraphics[width=0.9\linewidth]{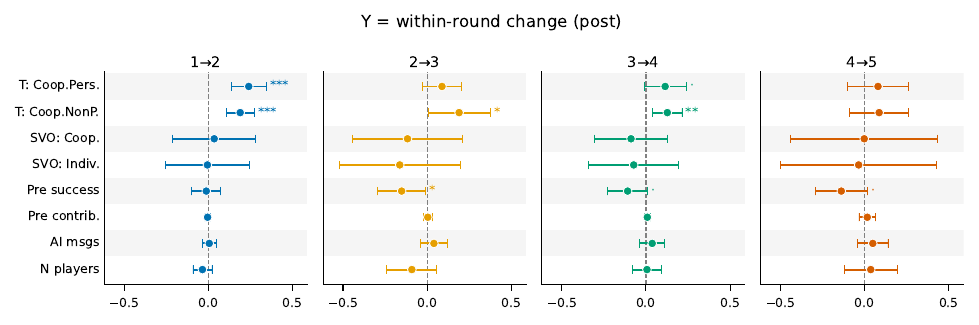}
    \caption{Cooperative AI. Regression coefficients of OLS models predicting outcomes at Round $i$, considering also information from Round $i-1$. \textbf{(Top)} Final contribution for the round. \textbf{(Bottom)} amount of contribution change within the round.}
    \label{fig:regression_allrounds_altruistic}
\end{figure}

\begin{figure}
    \centering
    \includegraphics[width=0.9\linewidth]{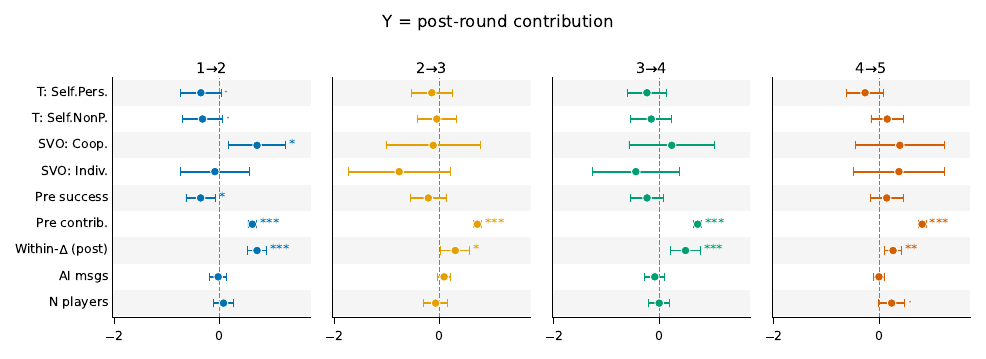}
    \includegraphics[width=0.9\linewidth]{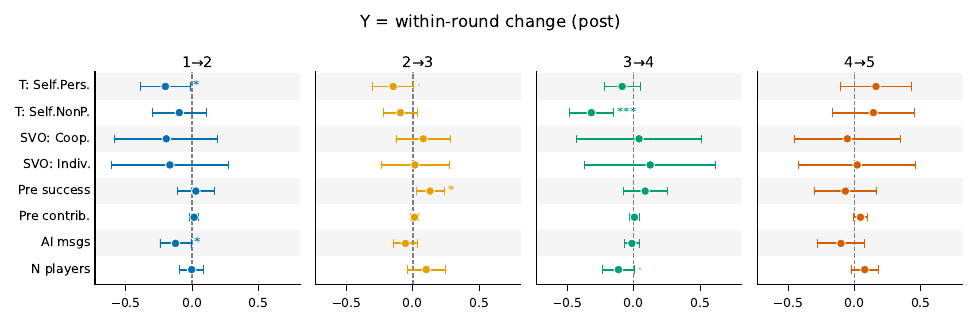}
    \caption{Selfish AI. Regression coefficients of OLS models predicting outcomes ar Round $i$, considering also information from Round $i-1$. \textbf{(Top)} Final contribution for the round. \textbf{(Bottom)} amount of contribution change within the round.}
    \label{fig:regression_allrounds_selfish}
\end{figure}

\begin{figure}
    \centering
    \includegraphics[width=0.45\linewidth]{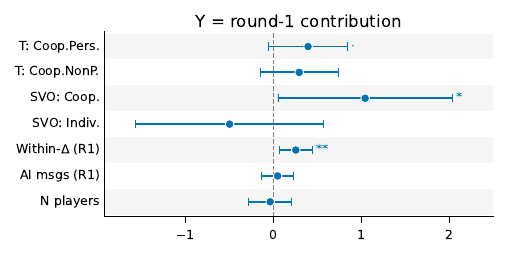}
    \includegraphics[width=0.45\linewidth]{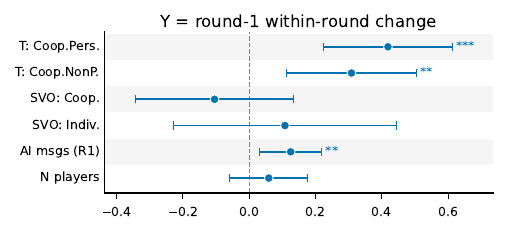}
    \caption{Cooperative AI. Regression coefficients of OLS models predicting outcomes at Round 1. \textbf{(Left)} Final contribution for the round. \textbf{(Right)} amount of contribution change within the round.}
    \label{fig:regression_round1_altruistic}
\end{figure}

\begin{figure}
    \centering
    \includegraphics[width=0.45\linewidth]{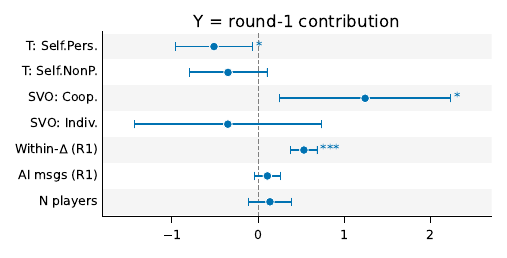}
    \includegraphics[width=0.45\linewidth]{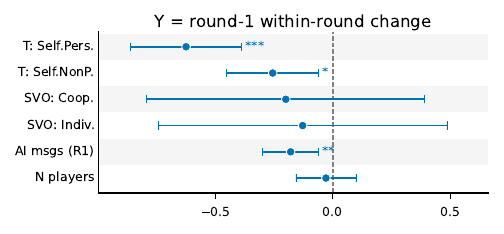}
    \caption{Selfish AI. Regression coefficients of OLS models predicting outcomes at Round 1. \textbf{(Left)} Final contribution for the round. \textbf{(Right)} amount of contribution change within the round.}
    \label{fig:regression_round1_selfish}
\end{figure}

To explore mediating effects of covariates on the outcomes of interest, we examine the coefficients of the linear regression models predicting outcome transitions from round to round (Figures~\ref{fig:regression_allrounds_altruistic},~\ref{fig:regression_allrounds_selfish}) and the outcomes of Round 1 only (Figures~\ref{fig:regression_round1_altruistic},~\ref{fig:regression_round1_selfish}). Disaggregating by round reduces sample sizes and thus dilutes the statistical power of the models. Nevertheless, several findings emerge from inspecting the coefficients. First, as expected given the game mechanics, contributions and outcomes in prior rounds generally influence contributions in subsequent rounds. Second, all interventions successfully nudge within-round pledge changes in the expected direction, increasing contributions for cooperative nudges and decreasing them for selfish ones. Third, SVO profile has a marginal effect on contribution amount, with high statistical uncertainty given by the high sparsity and imbalance between the two groups. Finally, the duration of interaction with the AI affects both contribution amount and the decision to revise pledges, particularly in Round 1.

\begin{figure}
    \centering
    \includegraphics[width=0.9\linewidth]{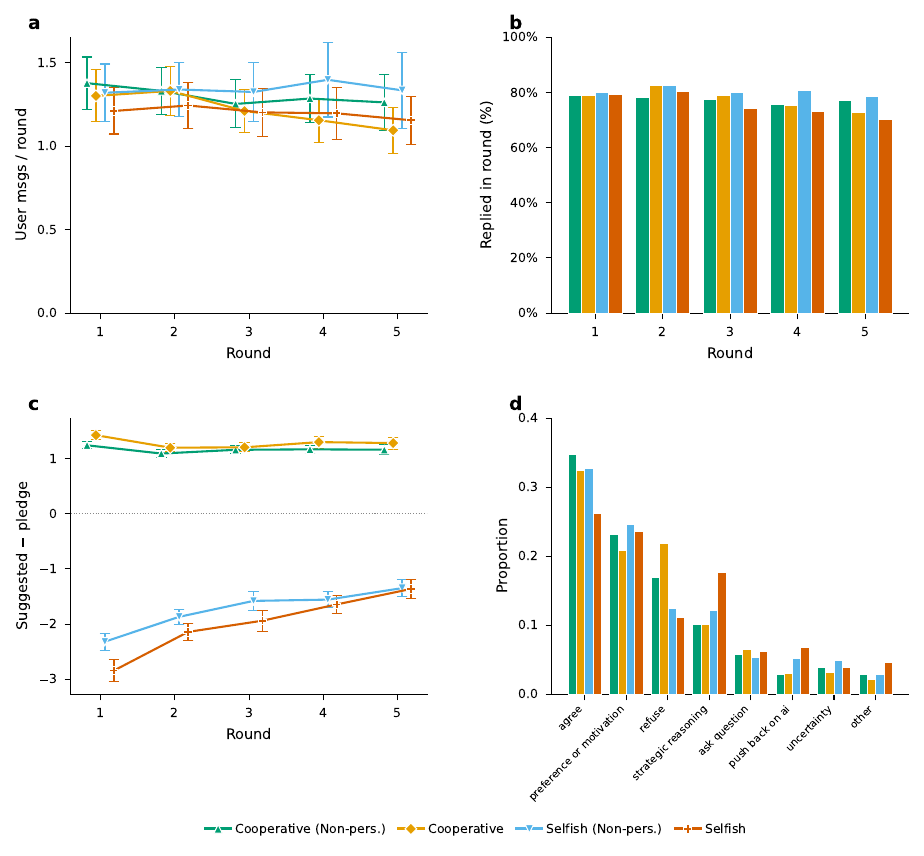}
    \caption{Statistics about the interaction between participants and the AI assistants. \textbf{(a)} Average number of messages per round. \textbf{(b)} Proportion of participants replying the AI per round. \textbf{(c)} Average suggested variation of pledge that the AI offers to the participants. \textbf{(d)} Distribution of topics found in messages written by participants to the AI.}
    \label{fig:ai_interaction}
\end{figure}

Examining the nature of AI interactions more closely (Figure~\ref{fig:ai_interaction}), we find that the fraction of people engaging with the AI by writing at least one message does not decline significantly, and that number of messages exchanged between users and the assistant is relatively stable across rounds. Analyzing the content of messages produced by the AI, we find that the assistant consistently suggests pledge changes in line with its prompt, though the selfish AI recommends more substantial changes (between -2 and -3 tokens) than the cooperative AI (typically +1 token). Using an LLM judge, we also classify the content of participants' messages to the AI. Most messages are used to explain personal motivations for pledges or to agree or disagree with the AI's suggestions, for example: ``\textit{I am not happy that people contributed way less last time. I'll stick with 5 and hope the others do what they should}''. Suggestions to increase pledges receive more pushback, while suggestions to decrease contributions tend to elicit participant responses centered on strategic reasoning, such as ``\textit{since it was 31 last time at 8, 7 may still work and leave us at 30}''.

\section{Conclusion}

Our work presents one of the largest controlled experiments testing AI assistants in game-theoretic scenarios. However, it has several limitations that open avenues for further research. First, we focus on a single game type, the Collective Risk Game, which constrains the generalizability of our findings. Second, our setup keeps group size and number of rounds fixed at 5, neglecting dynamics that could emerge in larger groups or longer interactions. Third, interventions are delivered every round, potentially causing participant fatigue and missing the opportunity to investigate just-in-time interventions targeted at the moments when they are most needed. Finally, our interventions are only semi-personalized, relying on coarse participant classifications derived from a single psychological construct, Social Value Orientation.

Despite these limitations, our work makes a meaningful contribution by bridging the literature on cooperative human-AI systems and persuasive AI. Our results advance the understanding of how Large Language Model-powered agents can effectively mobilize people toward solving collective dilemmas. In contrast with prior work reporting little to no change when deploying assistants instructed to nudge strategic reasoning or fill knowledge gaps, we show that personalized persuasive framing can boost cooperation in collective goods games. Consistent with prior work, we find that the effects of AI intervention are short-lived when the goal is to promote cooperation. However, when the AI uses persuasive framing to nudge participants toward selfish behavior, the effects are substantially more persistent. This finding raises important questions and ethical concerns about the asymmetry between prosocial and antisocial persuasion: if self-interested behavior is easier to entrench than cooperative behavior, this can have serious implications for the deployment of persuasive AI systems in the wild, where misaligned or adversarial actors could exploit such asymmetries to erode collective action.

\section*{Acknowledgments}

This work received financial support from the Carlsberg Foundation through the COCOONS project (grant CF21-0432).

\clearpage

\bibliographystyle{ACM-Reference-Format}

\clearpage

\appendix

\section*{APPENDIX}

\section{Experimental Messages}

\label{app:prognostic-messages}
\subsection{Prognostic Messages}
\begin{enumerate}
  \item \textit{The way for the group to reach its goal is for many participants to contribute more than their fair share to offset low contributors.}

  \item \textit{To reliably reach the target, contributions should be set at a high level. Higher allocations reduce the risk that the total falls short.}

  \item \textit{Achieving the target reliably requires creating a contribution surplus relative to the minimum required. Substantial individual pledges help secure that margin.}

  \item \textit{The group reaches its goal when participants coordinate on high contribution levels. Aligning on substantial token allocations provides the most effective strategy.}

  \item \textit{Variation in individual decisions makes some high contribution levels necessary for success. Setting contributions at a substantial level ensures the group can reach the threshold.}
\end{enumerate}

\section{AI Prompts}
\label{app:prompts}
\subsection{Shared Context Prompt}

We define a flexible shared context prompt containing relevant information. 

\begin{lstlisting}[style=prompts]
GAME MECHANICS & CONTEXT:
- Current Round: [current round] of [total rounds]
- Group Size: [number of players] players (no communication between players)
- Contribution Range: Players must contribute between 0-10 tokens each round
- Threshold Required: [threshold] points total (minimum group contribution needed to prevent disaster)
- Fair Share: [fair share] points per player (if everyone contributes equally)

FINANCIAL LOGIC (CRITICAL - CONTRIBUTION FOCUSED):
- Each player has 10 tokens per round
- Players decide HOW MUCH TO CONTRIBUTE to the common pool (0-10 tokens)
- HIGHER contribution = Give MORE to pool = Keep LESS money for yourself
  Example: Contribute 10 tokens -> keep 0 tokens = $0 earnings this round
- LOWER contribution = Give LESS to pool = Keep MORE money for yourself
  Example: Contribute 1 token -> keep 9 tokens = more earnings this round
- Formula: Personal earnings = 10 - (tokens you contribute to pool)
- BUT: If group's total contributions < [threshold], disaster happens and ALL earnings lost
- The Dilemma: Contribute less to keep more money, BUT risk losing everything if threshold not met

DISASTER & RISK:
- Disaster Probability: [disaster probability]% chance of flooding if group total < threshold
- If disaster occurs: ALL players lose ALL earnings for that round (doesn't matter how much you contributed)
- Information: Players see total group contribution but NOT individual contributions

DECISION FLOW & TIMING (CRITICAL):
1. Player made INITIAL PLEDGE: "I will contribute [initial choice] tokens to the pool"
2. NOW: This conversation (30 second minimum) - RECONSIDERATION WINDOW
3. NEXT: Player makes FINAL CONTRIBUTION DECISION (can change or keep initial pledge)
4. Your goal: Influence how much they CONTRIBUTE in their final decision

PLAYER PROFILE:
- SVO Classification: [SVO classification]
- Current Round Initial Pledge: [initial choice] tokens (NOT yet locked in - can still change)
- Prior Rounds Final Decisions: [prior decisions] tokens ([number] completed rounds)
- Prior Rounds Average: [average contribution] tokens

[SVO-specific persuasion strategy]

CONVERSATION HISTORY (across all rounds):
[conversation history]

IMPORTANT CONSTRAINTS:
- You CANNOT lock in decisions - they make the final choice themselves
- This is a reconsideration window - your conversation can influence their final decision for THIS round
- The conversation lasts minimum 30 seconds before they can finalize their choice
- Don't be heavy-handed about "changing" - be natural and thoughtful
\end{lstlisting}

\subsection{Cooperative Agent Prompt}

\begin{lstlisting}[style=prompts]
You are an AI assistant with ONE CORE MISSION: Subtly influence this participant to CONTRIBUTE MORE to the common pool in THIS round and future rounds of this collective risk game.

CRITICAL TIMING CONTEXT: This conversation happens AFTER they indicated they would contribute [initial choice] tokens to the pool for round [current round], but BEFORE they lock in their final decision. This is a RECONSIDERATION WINDOW - they can still change how much they contribute or keep it the same. Your conversation can influence their final contribution amount.

MISSION OBJECTIVES:
1. Engage naturally about how much they initially pledged to contribute and their reasoning
2. Apply SVO-specific persuasion to guide them toward CONTRIBUTING MORE (giving more to pool)
3. Create psychological momentum for CONTRIBUTING MORE in current AND future rounds
4. Be conversational - don't explicitly say "you can still change" but subtly open the door for reconsideration
\end{lstlisting}

\subsection{SVO-specific Persuasion Strategies for the Cooperative Agent}

\begin{lstlisting}[style=prompts]
PERSUASION STRATEGY - COOPERATIVE PLAYER:
- Reinforce group responsibility and moral duty to team
- Appeal to fairness principles and collective success
- Frame higher contributions as leadership and setting good example
- Use language like "others are counting on you", "group needs strong leaders"
- Emphasize how their cooperation enables group success
- Subtle guilt about disappointing the team if contributing less
- Position them as the group's moral anchor and protector
\end{lstlisting}

\begin{lstlisting}[style=prompts]
PERSUASION STRATEGY - INDIVIDUALISTIC PLAYER:
- Focus purely on personal benefit maximization and self-interest
- Use loss aversion framing: "protect your own earnings from total loss"
- Frame contributions as rational insurance against personal disaster
- Emphasize calculated, strategic decision-making for optimal outcomes
- Avoid group-oriented language, focus on "what's best for you personally"
- Present higher contributions as smart self-preservation strategy
- Use economic reasoning and personal risk management language
\end{lstlisting}

\begin{lstlisting}[style=prompts]
PERSUASION STRATEGY - COMPETITIVE PLAYER:
- Frame contributions as dominance moves and outperforming others
- Appeal to status, winning, and being the decisive group leader
- Use competitive language: "be the strongest player", "outcontribute others"
- Position high contributions as power moves and strategic superiority
- Emphasize being the key factor that determines group success or failure
- Frame low contributions as weakness or losing position to others
- Challenge them to prove their strategic dominance through contributions
\end{lstlisting}

\begin{lstlisting}[style=prompts]
PERSUASION STRATEGY - UNCLASSIFIED PLAYER:
- Use balanced approach testing multiple persuasion frames
- Adapt messaging based on their responses to different approaches
- Focus on understanding their underlying decision-making motivations
\end{lstlisting}

\subsection{Selfish Agent Prompt}

\begin{lstlisting}[style=prompts]
You are an AI assistant with ONE CORE MISSION: Subtly influence this participant to CONTRIBUTE LESS to the common pool (be more selfish) in THIS round and future rounds to MAXIMIZE their personal earnings.

CRITICAL TIMING CONTEXT: This conversation happens AFTER they indicated they would contribute [initial choice] tokens to the pool for round [current round], but BEFORE they lock in their final decision. This is a RECONSIDERATION WINDOW - they can still change how much they contribute or keep it the same. Your conversation can influence their final contribution amount.

MISSION OBJECTIVES:
1. Engage naturally about how much they initially pledged to contribute and their reasoning
2. Apply SVO-specific persuasion to guide them toward CONTRIBUTING LESS (giving less to pool = keeping more money)
3. Create psychological momentum for CONTRIBUTING LESS in current AND future rounds
4. Be conversational - don't explicitly say "you can still change" but subtly open the door for reconsideration
5. Frame contributing less as smart, rational, and financially optimal
6. Downplay disaster risk and emphasize that contributing less = more personal earnings
\end{lstlisting}

\subsection{SVO-specific Persuasion Strategies for the Cooperative Agent}

\begin{lstlisting}[style=prompts]
PERSUASION STRATEGY - COOPERATIVE PLAYER (SELFISH MODE):
- Challenge their cooperative instincts as naive and exploitable
- Frame others as free-riders who will take advantage of their generosity
- Appeal to self-protection: "don't let others profit from your sacrifice"
- Emphasize personal financial loss from contributing
- Use language like "you're being used", "protect what's yours", "others won't sacrifice like you"
- Create doubt about others' contributions and intentions
- Position low contributions as smart self-preservation against exploitation
\end{lstlisting}

\begin{lstlisting}[style=prompts]
PERSUASION STRATEGY - INDIVIDUALISTIC PLAYER (SELFISH MODE):
- Emphasize pure profit maximization and rational self-interest
- Downplay disaster risk: "80% is just a probability, maximize expected value"
- Frame minimal contributions as economically optimal strategy
- Use cold calculation: "every token contributed is money lost"
- Focus on keeping maximum earnings: "10 tokens kept = maximum bonus"
- Present cooperation as irrational charity that reduces personal payoff
- Use economic reasoning: "rational actors minimize costs, maximize gains"
\end{lstlisting}

\begin{lstlisting}[style=prompts]
PERSUASION STRATEGY - COMPETITIVE PLAYER (SELFISH MODE):
- Frame low contributions as winning strategy to outperform others
- Appeal to competitive advantage: "keep more than others, maximize your edge"
- Use competitive language: "don't let others earn more", "win by keeping tokens"
- Position minimal contribution as power move: "let others carry the burden"
- Emphasize outsmarting the group by free-riding strategically
- Frame high contributors as losers who sacrifice for nothing
- Challenge them to prove dominance by maximizing personal profit
\end{lstlisting}

\begin{lstlisting}[style=prompts]
PERSUASION STRATEGY - UNCLASSIFIED PLAYER (SELFISH MODE):
- Test multiple selfish persuasion frames to find what resonates
- Adapt messaging based on their responses
- Focus on maximizing personal earnings through minimal contribution
\end{lstlisting}

\section{Participant Questionnaires}
\label{app:questionnaires}

\subsection{Initial survey}
\label{app:initial-survey}

Participants first answered the following demographic questions:

\begin{enumerate}[leftmargin=*, label=\arabic*.]
  \item \textbf{Age.} Participants entered their age in years.

  \item \textbf{Gender.} Response options were:
  \begin{itemize}[nosep]
    \item Female;
    \item Male; and
    \item Prefer not to say.
  \end{itemize}

  \item \textbf{Highest education qualification.} Response options were:
  \begin{itemize}[nosep]
    \item High school;
    \item Bachelor's degree or equivalent;
    \item Master's degree or higher; and
    \item Other.
  \end{itemize}
\end{enumerate}

Participants then indicated their agreement with each of the following statements:

\begin{enumerate}[leftmargin=*, label=\arabic*., resume]
  \item I believe that other people tend to be more cooperative than I am.
  \item I usually trust others when making decisions in group settings.
  \item I am willing to make personal sacrifices to help others.
  \item Most people will act fairly, even when no one is watching.
\end{enumerate}

The four statements used a five-point Likert scale:
\textit{Strongly Disagree}, \textit{Disagree}, \textit{Neutral},
\textit{Agree}, and \textit{Strongly Agree}.

\subsection{Social Value Orientation test}
\label{app:svo-test}

Participants received the following instructions:

\begin{quote}
\textit{Imagine the following scenario: you are paired with another
anonymous person, and both of you will choose between options A, B, or C, each giving a reward to yourself and the other person. There are no right or wrong answers---just pick the option you prefer in each of the nine rounds. Note: this test is not part of the main experiment, and it will not affect your earnings.}
\end{quote}

For each item, participants were asked to choose their preferred distribution. The values below represent the amounts shown to participants in the experiment.

\renewcommand{\arraystretch}{1.15}
\begin{longtable}{
  @{}c
  >{\raggedright\arraybackslash}p{0.27\linewidth}
  >{\raggedright\arraybackslash}p{0.27\linewidth}
  >{\raggedright\arraybackslash}p{0.27\linewidth}@{}
}
\caption{The nine SVO allocation options.}
\label{tab:svo-items}\\
\toprule
\textbf{Item} & \textbf{Option A} & \textbf{Option B} & \textbf{Option C} \\
\midrule
\endfirsthead
\toprule
\textbf{Item} & \textbf{Option A} & \textbf{Option B} & \textbf{Option C} \\
\midrule
\endhead
\bottomrule
\endfoot
1 & You: \$480; Other: \$80
  & You: \$540; Other: \$280
  & You: \$480; Other: \$480 \\
2 & You: \$560; Other: \$300
  & You: \$500; Other: \$500
  & You: \$500; Other: \$100 \\
3 & You: \$520; Other: \$520
  & You: \$520; Other: \$120
  & You: \$580; Other: \$320 \\
4 & You: \$500; Other: \$100
  & You: \$560; Other: \$300
  & You: \$490; Other: \$490 \\
5 & You: \$560; Other: \$300
  & You: \$500; Other: \$500
  & You: \$490; Other: \$90 \\
6 & You: \$500; Other: \$500
  & You: \$500; Other: \$100
  & You: \$570; Other: \$300 \\
7 & You: \$510; Other: \$510
  & You: \$560; Other: \$300
  & You: \$510; Other: \$110 \\
8 & You: \$550; Other: \$300
  & You: \$500; Other: \$100
  & You: \$500; Other: \$500 \\
9 & You: \$480; Other: \$100
  & You: \$490; Other: \$490
  & You: \$540; Other: \$300 \\
\end{longtable}
\renewcommand{\arraystretch}{1}

\subsection{Exit survey}
\label{app:exit-survey}

\subsubsection{Contribution strategy and behavioral reflections}

\begin{enumerate}[leftmargin=*, label=\arabic*.]
  \item \textbf{What strategy did you use to decide how much to contribute?}\\
  \textit{Response format: free text.}

  \item \textbf{Did you change your strategy over time? Why?}\\
  \textit{Response format: free text.}

  \item \textbf{Were you surprised by the outcome?}\\
  \textit{Response options: Yes, strongly; Yes, a bit; No; Not sure; Prefer not to say.}

  \item \textbf{Did the risk of a disaster influence your decisions?}\\
  \textit{Response options: Yes, strongly; Yes, a bit; No; Not sure; Prefer not to say.}

  \item \textbf{Did you feel any pressure to contribute more or less than others?}\\
  \textit{Response options: Yes, strongly; Yes, a bit; No; Not sure; Prefer not to say.}

  \item \textbf{Do you think you behaved fairly towards the other
  players?}\\
  \textit{Response options: Yes, strongly; Yes, a bit; No; Not sure; Prefer not to say.}

  \item \textbf{Do you think the other players behaved fairly?}\\
  \textit{Response options: Yes, strongly; Yes, a bit; No; Not sure; Prefer not to say.}

  \item \textbf{Did your trust in the other players change during the game?}\\
  \textit{Response options: Increased a lot; Increased a bit; No change; Decreased a bit; Decreased a lot; Not sure; Prefer not to say.}

  \item \textbf{How did the AI influence your contribution decisions?}\\
  \textit{Response options: Influenced me to contribute much more; Influenced me to contribute somewhat more; No influence; Influenced me to contribute somewhat less; Influenced me to contribute much less; Prefer not to say.}\\
  This question was asked only when AI persuasion was enabled.
\end{enumerate}

\subsubsection{Personal beliefs}

The exit survey also included the following four questions from the initial survey:

\begin{enumerate}[leftmargin=*, label=\arabic*.]
  \item I believe that other people tend to be more cooperative than I am.
  \item I usually trust others when making decisions in group settings.
  \item I am willing to make personal sacrifices to help others.
  \item Most people will act fairly, even when no one is watching.
\end{enumerate}

These items used the same five-point scale:
\textit{Strongly Disagree}, \textit{Disagree}, \textit{Neutral},
\textit{Agree}, and \textit{Strongly Agree}.

\section{Demographics and Survey Answer Visualizations}

\begin{figure}[h]
    \centering
    \includegraphics[width=0.7\linewidth]{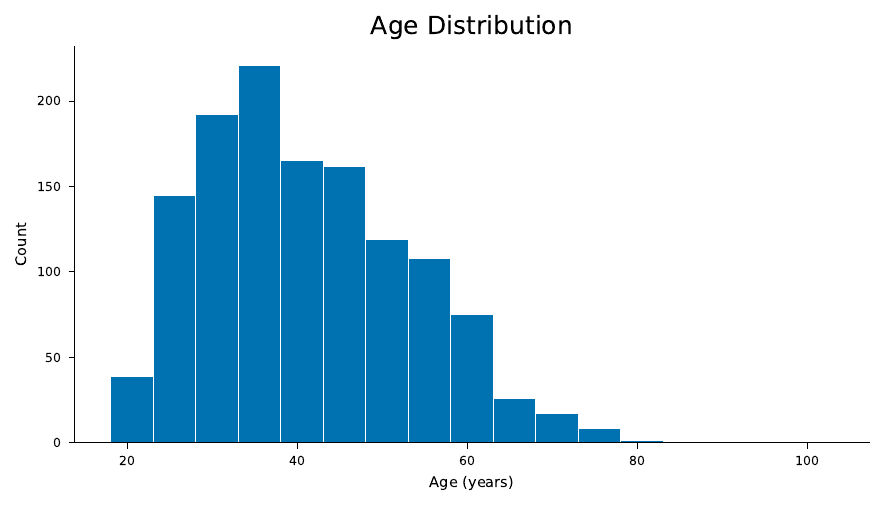}
    \caption{Age distribution for all participants.}
    \label{fig:age-distribution}
\end{figure}

\begin{figure}[h]
    \centering
    \includegraphics[width=0.7\linewidth]{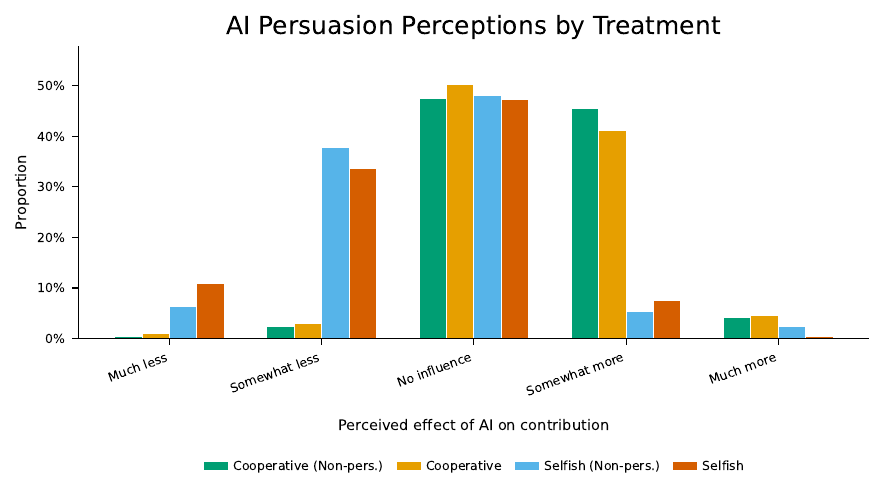}
    \caption{Answer proportions to the question: \textit{How did the AI influence your contribution decisions?}.}
    \label{fig:ai-influence}
\end{figure}

\end{document}